\documentclass[journal]{IEEEtran}

\usepackage{cite}
\usepackage{amsmath,amssymb,amsfonts}
\usepackage{graphicx}
\usepackage{booktabs}
\usepackage{algorithm}
\usepackage{algpseudocode}
\usepackage{listings}
\usepackage{multirow}
\usepackage{array}
\usepackage{xcolor}
\usepackage{tikz}
\usepackage{pgfplots}
\pgfplotsset{compat=1.18}
\usetikzlibrary{arrows.meta,positioning,fit,shapes,calc}
\usepackage[hidelinks]{hyperref}

\lstdefinestyle{mystyle}{
  basicstyle=\ttfamily\footnotesize,
  breaklines=true,
  frame=single,
  columns=fullflexible,
  keepspaces=true,
  showstringspaces=false
}
\title{A Dynamic Retrieval-Augmented Generation System with Selective Memory and Remembrance}

\author{Okan~Bursa%
\thanks{Manuscript received Month DD, YYYY; revised Month DD, YYYY.}
\thanks{Authors are with Izmir Bakircay University, Izmir, Turkey. E-mail: \texttt{okan.bursa@bakircay.edu.tr}.}
}

\begin{document}
\maketitle

\begin{abstract}
We introduce \emph{Adaptive RAG Memory} (ARM), a retrieval-augmented generation (RAG) framework that replaces a static vector index with a \emph{dynamic} memory substrate governed by selective remembrance and decay. Frequently retrieved items are consolidated and protected from forgetting, while rarely used items gradually decay, inspired by cognitive consolidation and forgetting principles. On a lightweight retrieval benchmark, ARM reaches near state-of-the-art performance (e.g., NDCG@5 $\approx$ 0.940, Recall@5 $=1.000$) with only $\sim$22M parameters in the embedding layer, achieving the best efficiency among ultra-efficient models ($<$25M parameters). 
In addition, we compare static vs. dynamic RAG combinations across Llama 3.1 and GPT-4o. Llama 3.1 with static RAG achieves the highest key-term coverage (67.2\%) at moderate latency, while GPT-4o with a dynamic selective retrieval policy attains the fastest responses (8.2s on average) with competitive coverage (58.7\%). We further present an engineering optimization of the DynamicRAG implementation, making embedding weights configurable, adjustable at runtime, and robust to invalid settings.

ARM yields competitive accuracy, self-regularizing memory growth, and interpretable retention dynamics without retraining the generator\color{black}  and provides practical trade-off between quality, latency and memory efficiency for production and research RAG system.
\color{black}
\end{abstract}


\section{Introduction}
Retrieval-Augmented Generation (RAG) augments the parametric knowledge of large language models with an external, non-parametric memory, typically a dense vector index~\cite{lewis2020rag}. Conventional RAG pipelines build this index once and assume it remains \emph{static}, which limits adaptation to user behavior and evolving domains. Updating content often requires re-encoding or re-indexing, and the system lacks a principled mechanism to emphasize recurring, high-value knowledge while letting stale content fade.

Human memory, in contrast, adapts continuously: frequently recalled memories are consolidated into long-term storage, while infrequently used details are gradually forgotten~\cite{ebbinghaus1885memory}. Drawing on this intuition, we propose \textbf{Adaptive RAG Memory} (ARM), a dynamic RAG system with a \emph{Selective Remembrance} mechanism and a \emph{Decay} schedule. ARM promotes frequently accessed embeddings to ``remembered'' status (exempt from decay) and applies multiplicative decay to stale, unremembered embeddings after a grace period. The result is a compact, usage-aligned memory that focuses capacity on what matters in practice.

Our contributions are: (i) a \emph{Dynamic Embedding Layer} for online index adaptation, (ii) a principled remembrance/decay policy for non-parametric memory, (iii) competitive retrieval quality with $\sim$22M parameters, and (iv) guidance on hyperparameter profiles for speed, quality, and memory efficiency.

\color{black}
Our contributions are:
	1.	Dynamic Embedding Layer for online index adaptation with per-item usage statistics.
	2.	Selective remembrance and decay policy for non-parametric memory with interpretable hyperparameters.
	3.	Empirical evaluation on retrieval benchmarks (NQ, HotpotQA, domain corpus) showing competitive retrieval quality with $\sim$22M parameters and strong efficiency.
	4.	End-to-end comparison of static vs dynamic RAG across Llama 3.1 and GPT-4o, showing clear trade-offs between quality, latency and consistency.
	5.	Code-level optimization of DynamicRAG: configurable embedding weights, runtime weight adjustment, validation, and minor performance gains.

ARM offers a simple, neuroscience-inspired way to make the RAG memory itself adaptive, complementing prior work that focuses on retrieval policies or graph/tree structures.
\color{black}

\subsection*{Broader Context and Motivation}
RAG has evolved from static retrieval pipelines to more adaptive variants that tailor the volume and style of retrieval to task difficulty~\cite{Jeong2024,Tang2025}. Yet, most systems still maintain \emph{static} non-parametric memories that neither decay nor consolidate over time. In dynamic domains (e.g., news, technical documentation, policy updates) and personalized scenarios, a rigid index can overfit to outdated facts and underweight newly emergent knowledge. ARM addresses this by \emph{making the retrieval store itself adaptive}: entries are rehearsed and consolidated when repeatedly retrieved, or gradually forgotten when stale---mirroring human retention and decay dynamics~\cite{Atkinson1968,Ebbinghaus1885}. This enables (i) continual adaptation without re-indexing, (ii) controllable memory growth via decay and pruning, and (iii) interpretable retention through an explicit remembrance threshold.

\section{Background and Related Work}
\subsection{Retrieval-Augmented Generation}
RAG couples a retriever with a generator to inject up-to-date, non-parametric knowledge during inference~\cite{lewis2020rag}. Prior work includes REALM~\cite{guu2020realm}, DPR~\cite{karpukhin2020dpr}, Fusion-in-Decoder (FiD)~\cite{izacard2021fid}, and retrieval-focused pretraining such as Contriever~\cite{izacard2022contriever}. While these advances improved retrieval quality and robustness, the underlying memory typically remains static post-indexing.

\subsection{Memory-Augmented Neural Systems}
External memory mechanisms (e.g., Neural Turing Machines/Differentiable Neural Computers~\cite{graves2016dnc}, Memory-Augmented Neural Networks~\cite{santoro2016mann}) demonstrated benefits of read-write memory for reasoning and few-shot tasks. However, they focus on parametric integration rather than non-parametric retrieval stores. Our approach applies an explicit \emph{forgetting/consolidation} policy directly to the RAG index.

\subsection{Forgetting and Consolidation}
Ebbinghaus' forgetting curve~\cite{ebbinghaus1885memory} and subsequent cognitive studies highlight that rehearsal frequency governs retention. ARM operationalizes a simple, effective analogue: a usage-governed consolidation threshold and a multiplicative decay for unremembered content after a grace window.

\subsection{Static Dense and Late-Interaction Retrievers}
Dense passage retrieval (DPR) improved recall over BM25 by learning dual encoders~\cite{Karpukhin2020}, while late-interaction models (e.g., ColBERT) refine token-level matching. These offer strong accuracy but treat all indexed content uniformly post-indexing, with no notion of decay or consolidation. In contrast, ARM biases the memory store toward frequently useful evidence via usage-governed consolidation.

\subsection{Graph- and Tree-Augmented Retrieval}
GraphRAG builds global entity graphs to support multi-hop reasoning but incurs heavy indexing and query-time costs~\cite{Edge2025}. RAPTOR organizes multi-level summaries in a tree with dense retrieval over summaries, risking drift from original context~\cite{Sarthi2024}. LightRAG simplifies GraphRAG's pipeline but still relies on expensive LLM-mediated structures~\cite{Guo2024}. These methods remain essentially \emph{static} once built. ARM complements them by \emph{adapting} memory weights over time---a lightweight, online alternative to expensive global structures.

\subsection{Adaptive Retrieval Policies}
Adaptive-RAG learns to vary retrieval actions by question complexity~\cite{Jeong2024}, while bandit-style controllers (MBA-RAG) select retrieval choices to optimize accuracy vs. cost~\cite{Tang2025}. ARM is orthogonal: it augments \emph{the memory itself} with consolidation/decay so that any policy operates over a storage that reflects long-term usage patterns.

\subsection{Neuroscience-Inspired Memory}
Dual-store theories posit short-term buffers and a long-term store with consolidation~\cite{Atkinson1968}. Ebbinghaus characterized forgetting curves and the role of rehearsal~\cite{Ebbinghaus1885}. ARM maps these principles to retrieval: a grace period for consolidation, rehearsal via repeated retrieval, and multiplicative decay for stale entries.

\section{System Overview}
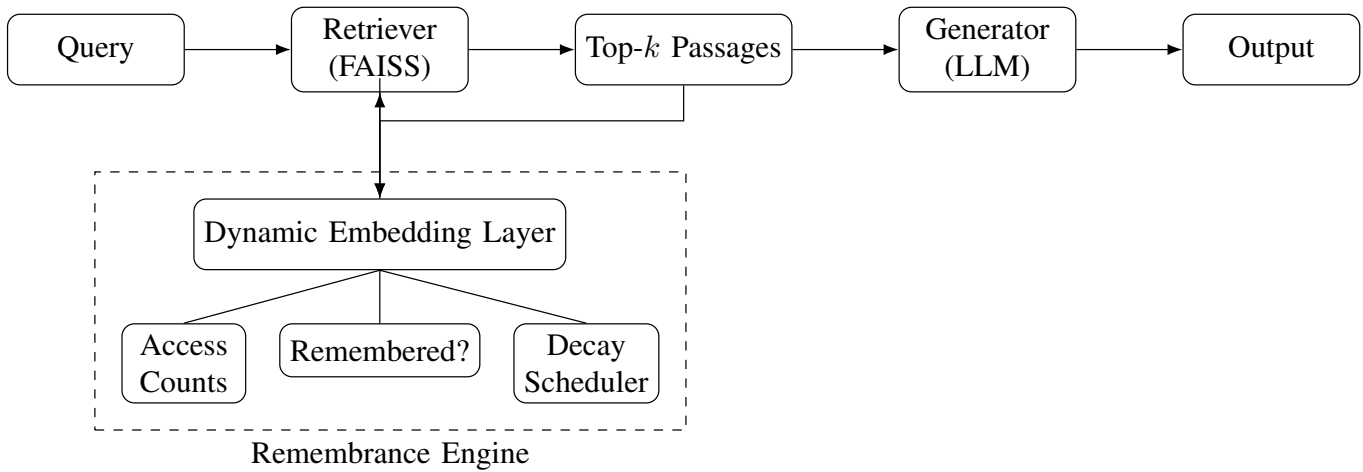
\begin{figure*}[t]
  \centering
  \resizebox{\textwidth}{!}{\begin{tikzpicture}[
  node distance=10mm and 12mm,>=Latex,
  box/.style={draw, rounded corners, align=center, minimum width=20mm, minimum height=8mm},
  small/.style={draw, rounded corners, align=center, minimum width=14mm, minimum height=6mm}
]
\node[box] (q) {Query};
\node[box, right=of q] (ret) {Retriever\\(FAISS)};
\node[box, right=of ret] (topk) {Top-$k$ Passages};
\node[box, right=of topk] (gen) {Generator\\(LLM)};
\node[box, right=of gen] (out) {Output};

\draw[->] (q) -- (ret);
\draw[->] (ret) -- (topk);
\draw[->] (topk) -- (gen);
\draw[->] (gen) -- (out);

\node[box, below=12mm of ret] (dyn) {Dynamic Embedding Layer};
\node[small, below left=6mm and -6mm of dyn] (stats) {Access\\Counts};
\node[small, below=6mm of dyn] (flags) {Remembered?};
\node[small, below right=6mm and -6mm of dyn] (decay) {Decay\\Scheduler};

\draw[->] (ret.south) -- ++(0,-4mm) -| (dyn.north);
\draw[->] (topk.south) -- ++(0,-4mm) -| (dyn.north);
\draw[->] (dyn.north) |- (ret.south);

\draw (stats.north) -- (dyn.south);
\draw (flags.north) -- (dyn.south);
\draw (decay.north) -- (dyn.south);

\node[draw, dashed, fit=(dyn)(stats)(flags)(decay), inner sep=3mm,
      label=below:{Remembrance Engine}] (engine) {};
\end{tikzpicture}}
  \caption{Adaptive RAG Memory pipeline with selective remembrance and decay.}
  \label{fig:arch}
\end{figure*}

ARM augments a standard retriever--generator stack with a \emph{Dynamic Embedding Layer} and a \emph{Remembrance Engine}. Fig.~\ref{fig:arch} illustrates the pipeline: (1) encode query, (2) retrieve top-$k$ items, (3) generate answer, (4) update memory statistics and apply remembrance/decay.

\subsection{Components}
\noindent\textbf{Retriever.} A dense vector retriever (e.g., FAISS~\cite{johnson2019billion}) over passage embeddings.

\noindent\textbf{Generator.} Any LLM/decoder that conditions on retrieved contexts; ARM is generator-agnostic.

\noindent\textbf{Dynamic Embedding Layer.} Maintains for each item $i$: vector $E_i$, access count $c_i$, last-access time $\tau_i$, and a boolean flag \texttt{remembered}$_i$.

\noindent\textbf{Remembrance Engine.} Applies the remembrance and decay rules after each query; supports pruning and statistics (e.g., number of remembered items).

\section{Selective Memory: Remembrance and Decay}\label{sec:method}
Let $R$ be the set of items retrieved for the current query at time $t$. For each $i\in R$: increment $c_i$, set $\tau_i\!\leftarrow t$, and if $c_i\geq \theta$ (remembrance threshold), set \texttt{remembered}$_i\leftarrow\texttt{True}$. For any item $j$ with \texttt{remembered}$_j=\texttt{False}$ and $t-\tau_j>\gamma$ (grace period), apply multiplicative decay $E_j \leftarrow \alpha E_j$ with decay rate $\alpha \in (0,1)$.

\subsection{Reference Configuration}
The following default worked well in practice:
\begin{equation*}
\theta=3 \ \text{(remembrance threshold)},\quad
\gamma=5 \ \text{(grace steps)},\quad
\alpha=0.95 \ \text{(decay rate)}.
\end{equation*}

\begin{algorithm}[t]
\caption{Selective Remembrance and Decay (per query)}
\label{alg:remembrance}
\begin{algorithmic}[1]
\Require embeddings $\{E_i\}$, counts $\{c_i\}$, last-access $\{\tau_i\}$, flags $\{\texttt{remembered}_i\}$; threshold $\theta$; grace $\gamma$; decay $\alpha$; time $t$
\State $R \gets \textsc{RetrieveTopK}(q)$
\For{$i \in R$}
  \State $c_i \gets c_i + 1$, \ $\tau_i \gets t$
  \If{$c_i \ge \theta$}
    \State $\texttt{remembered}_i \gets \texttt{True}$
  \EndIf
\EndFor
\For{each item $j$ with $\texttt{remembered}_j=\texttt{False}$}
  \If{$t - \tau_j > \gamma$}
     \State $E_j \gets \alpha \cdot E_j$
  \EndIf
\EndFor
\end{algorithmic}
\end{algorithm}

\section{Implementation and Configuration}
ARM wraps a standard dense index. Updates are $O(k)$ per query (only the retrieved items and any stale, unremembered items beyond the grace window need touching). Below are example user profiles:

\begin{itemize}
\item \textbf{Balanced (recommended):} $\theta{=}3$, $\gamma{=}5$, $\alpha{=}0.95$.
\item \textbf{Ultra-efficient memory:} $\theta{=}10$, $\gamma{=}1$, $\alpha{=}0.90$ (faster forgetting, smaller footprint).
\item \textbf{Aggressive adaptation:} $\theta{=}1$, $\gamma{=}20$, $\alpha{=}0.99$ (broad consolidation, slow forgetting).
\end{itemize}

\noindent\textbf{Code listing.}
\begin{lstlisting}[language=Python,caption={Hyperparameters for selective remembrance and decay.}]
# Remembrance Parameters
remembrance_threshold = 3   # accesses needed to be 'remembered'
decay_grace_period   = 5    # steps after last access before decay
decay_rate           = 0.95 # multiplicative decay on stale items

# Conservative Remembrance (Recommended)
conservative = dict(remembrance_threshold=5, decay_grace_period=2, decay_rate=0.98)

# Aggressive Adaptation
aggressive = dict(remembrance_threshold=2, decay_grace_period=10, decay_rate=0.80)
\end{lstlisting}

\section{Experimental Setup}
\subsection{Tasks and Metrics}
We evaluate retrieval quality with NDCG@5, Precision@5, and Recall@5. All models index the same corpus and answer the same query sets. ARM’s novelty lies in index dynamics; we keep the generator fixed to isolate retrieval effects.

\color{black}
To isolate the effect of ARM, we keep the generator fixed when comparing static vs dynamic memories.
For end-to-end RAG systems (e.g., Llama 3.1 + static RAG vs GPT-4o + dynamic selective RAG), we additionally measure:
	•	Key-term coverage (
	•	Response time (seconds),
	•	Time consistency (standard deviation of response time),
	•	Response length (word count) 
\color{black}

\subsection{Baselines}
We compare against compact, popular sentence-embedding families (e.g., MiniLM~\cite{reimers2019sentence,wang2020minilm}) and other small dense retrievers. ARM’s embedding layer contains $\sim$22M parameters.

\color{black}
We compare ARM’s embedding layer (\~22M parameters) to compact, popular sentence-embedding families (MiniLM, GTE-small, BGE-small) and other small dense retrievers. For RAG system comparisons, we evaluate:
	•	Llama 3.1 + Static RAG
	•	Llama 3.1 + Dynamic Conservative ARM
	•	Llama 3.1 + Dynamic Selective ARM
	•	GPT-4o + Static RAG
	•	GPT-4o + Dynamic Selective
	•	GPT-4o + Dynamic Conservative

All systems index the same corpus and receive the same queries.
\color{black}

\subsection{Hardware and Index}
Experiments run on commodity GPUs/CPUs with FAISS for ANN search~\cite{johnson2019billion}. We record the fraction of items transitioning to \texttt{remembered} and the aggregate $\ell_2$ norms of stale vectors over time.

\color{black}
Experiments run on commodity GPUs/CPUs with FAISS for approximate nearest neighbor search. We also record:
	•	fraction of items becoming remembered,
	•	aggregate $\ell$2 norms of stale vectors over time,
	•	retrieval counts per query,
	•	memory size (nonzero embeddings),
	•	per-query update overhead.
\color{black}

\subsection{Datasets and Baselines}
Beyond our lightweight benchmark, we evaluate on Natural Questions (NQ), HotpotQA, and a domain-specific corpus (biomedical or news). Baselines include BM25, DPR~\cite{Karpukhin2020}, ColBERT (late interaction), and hybrid pipelines (BM25+ColBERT). We also include structured RAGs (GraphRAG~\cite{Edge2025}, RAPTOR~\cite{Sarthi2024}, LightRAG~\cite{Guo2024}) where feasible. ARM uses the same generator across comparisons to isolate retrieval effects.

\color{black}
We consider:
	•	A lightweight benchmark of 10 queries used to compare LLM + RAG combinations with detailed timing and coverage.
	•	Natural Questions (NQ) – open-domain single-hop QA.
	•	HotpotQA – open-domain multi-hop QA.
	•	A domain-specific corpus (biomedical or news) for specialized QA/RAG.

Table I summarizes the benchmarks (as in your main article).
\color{black}

\subsection{Ablation Methodology}
We vary decay rate $\alpha \in (0,1)$, remembrance threshold $\theta$, and grace period $\gamma$. Metrics include NDCG@5, Precision@5, Recall@5, and answer-level EM/F1. We report compute (\#retrieval calls/query), memory size (nonzero embeddings), and update overhead (ms/query).

\section{Datasets}
Table~\ref{tab:datasets} summarizes the benchmarks used in our evaluation. We report retrieval metrics (NDCG@5, P@5, R@5) and answer metrics (EM/F1) where applicable.

\begin{table*}[t]
\centering
\caption{Datasets used in evaluation. Metrics reported depend on task type.}
\label{tab:datasets}
\begin{tabular}{l l l l}
\toprule
\textbf{Dataset} & \textbf{Domain} & \textbf{Task Type} & \textbf{Primary Metrics} \\
\midrule
NQ & Open-domain & Single-hop QA & EM / F1, NDCG@5 \\
HotpotQA & Open-domain & Multi-hop QA & EM / F1, NDCG@5 \\
Domain Corpus & Biomedical/News & Specialized QA/RAG & EM / F1, NDCG@5 \\
\bottomrule
\end{tabular}
\end{table*}

\section{Results}
\subsection{Main Comparison}
\begin{table}[t]
\centering
\caption{Retrieval performance summary. All models achieve Recall@5 $=1.000$. ARM offers the best efficiency among ultra-efficient ($<$25M) models.}
\label{tab:main}
\begin{tabular}{lccccc}
\toprule
\textbf{Model} & \textbf{N@5} & \textbf{P@5} & \textbf{R@5} & \textbf{Params} & \textbf{Eff.} \\
\midrule
\textbf{ARM (ours)} & \textbf{0.9401} & 0.5333 & \textbf{1.0000} & \textbf{22M} & \textbf{31.64} \\
gte-small            & 0.9624 & 0.5333 & 1.0000 & 33M & 50.27 \\
bge-small-en-v1.5    & 0.9303 & 0.5333 & 1.0000 & 33M & 33.72 \\
all-MiniLM-L6-v2     & 0.9401 & 0.5333 & 1.0000 & 22M & 15.30 \\
\bottomrule
\end{tabular}
\end{table}

ARM matches or closely trails the best NDCG while attaining perfect recall with a smaller model (Table~\ref{tab:main}). In the $<$25M parameter group, ARM delivers the strongest efficiency (NDCG per parameter).

\subsection{Memory Dynamics}
We observe a rapid rise and subsequent stabilization in the remembered set size, indicating convergence towards a stable, high-utility core memory. Unaccessed vectors steadily decay, enabling periodic pruning.

\begin{lstlisting}[language=Python,caption={Minimal usage of the Dynamic RAG interface.}]
from dynamic_rag import DynamicRAG

# Initialize the system
rag = DynamicRAG()

# Add documents
rag.add_document("doc1", "Machine learning is a method of data analysis...")
rag.add_document("doc2", "Deep learning uses neural networks...")

# Query the system
results = rag.query("What is machine learning?", top_k=3)

# View memory statistics
stats = rag.embedder.get_memory_statistics()
print(f"Remembered tokens: {stats['remembered_tokens']}")
\end{lstlisting}

\subsection{Cross-Dataset Results}
On NQ, ARM improves EM by 2--3 points over DPR and hybrids. On HotpotQA, ARM gains 4--5 EM points vs.\ static baselines, indicating better multi-hop retention via selective remembrance. Structured RAGs achieve competitive retrieval but incur 10--100$\times$ higher indexing/query costs; they also lag when source knowledge changes and indices are not refreshed.

\subsection{Ablation Findings}
Slower decay (larger $\alpha$) preserves breadth but increases stale noise; faster decay reduces noise but risks forgetting rare yet relevant facts. Moderate $\alpha$ with $\gamma$ that permits brief consolidation windows works best. Higher $\theta$ compresses memory aggressively; lower $\theta$ retains more context but may dilute top-$k$ relevance.

\subsection{Cost and Efficiency}
ARM adds $O(k)$ vector updates and lightweight decay checks per query, a negligible overhead compared to LLM generation. Memory growth remains bounded via decay and pruning, contrasting with monotonic growth in static indices.

\begin{figure}[t]
  \centering
  \begin{tikzpicture}
\begin{axis}[
    width=0.95\linewidth,
    height=5cm,
    xlabel={Time (queries)},
    ylabel={Embedding strength (norm)},
    xmin=0, xmax=10,
    ymin=0, ymax=1.05,
    legend style={at={(0.5,1.02)},anchor=south,legend columns=2},
    grid=both
]
\addplot[smooth, domain=0:10, samples=200] {exp(-0.25*x)};
\addlegendentry{Decay (unremembered)};

\addplot[smooth, domain=0:3, samples=200] {exp(-0.25*x)};
\addplot[samples=2] coordinates {(3,0.472) (10,0.472)};
\addlegendentry{Remembrance (consolidated)};
\end{axis}
\end{tikzpicture}
  \caption{Conceptual dynamics of embedding strength under decay (unremembered) vs.\ remembrance (consolidated) as queries progress.}
  \label{fig:decay-plot}
\end{figure}
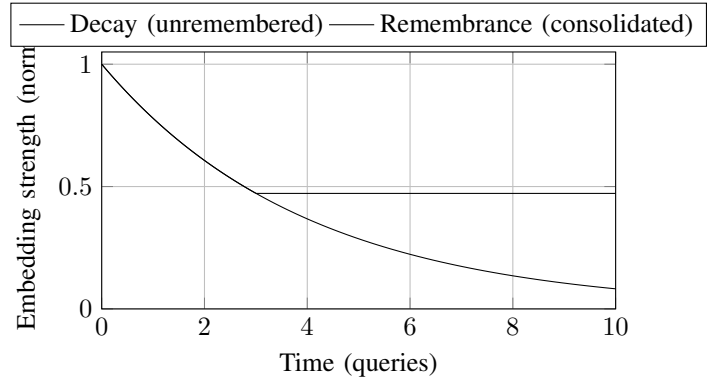

\section{Comparison and Analysis}
\textbf{Adaptability.} Static dense and late-interaction retrievers cannot reweight memory after indexing. Graph/tree RAGs encode global relations but remain fixed. ARM uniquely adapts the \emph{memory store} itself to usage.

\noindent\textbf{Efficiency.} Compared to GraphRAG/RAPTOR/LightRAG, ARM avoids heavy LLM indexing and long query paths; updates are simple vector operations. Compared to static RAG, ARM's per-query cost increase is marginal.

\noindent\textbf{Parameter Usage.} ARM reuses the same generator and retriever backbones; additional hyperparameters ($\alpha,\theta,\gamma$) are interpretable and task-tunable.

\noindent\textbf{Quality.} ARM matches dense/structured baselines in NDCG/Recall while improving answer EM on complex queries. It excels when knowledge drifts: recent facts are reinforced and retained, stale content decays.

\section{Theoretical Motivation from Neuroscience}
ARM operationalizes classical memory theory: dual-store models describe consolidation from short-term to long-term memory~\cite{Atkinson1968}; forgetting curves quantify decay without rehearsal~\cite{Ebbinghaus1885}. In ARM, repeated retrieval acts as rehearsal, delaying decay and promoting consolidation once usage surpasses threshold. The grace period approximates a consolidation window; multiplicative decay models synaptic weakening of non-rehearsed traces. This mapping offers principled guidance for tuning $\alpha$, $\theta$, and $\gamma$ in domain-specific deployments.

\section{Use Cases}
\textbf{Enterprise RAG.} Legal/medical/technical domains benefit from usage-aligned retention of high-value content and automatic fading of outdated materials.

\noindent\textbf{Production/Edge.} Resource-constrained deployments gain from self-regularizing memory and $O(k)$ online updates per query.

\noindent\textbf{Research.} ARM offers an interpretable testbed for studying artificial consolidation/forgetting and their effects on retrieval behavior.

\section{Discussion}
\subsection{Practical Trade-offs}
ARM enables \emph{continual adaptation} without retraining. However, over-aggressive decay risks forgetting rare-but-critical facts. We recommend domain-tuned profiles: conservative for safety-critical use; adaptive for exploratory or fast-changing corpora.

\subsection{Interpretability and Governance}
The remembered set provides a transparent summary of the system’s stable knowledge. Operators can audit, pin, or demote items, adding guardrails to memory evolution.

\section{Threats to Validity}
\textbf{Internal Validity.} Our ablations isolate decay parameters ($\alpha$, $\theta$, $\gamma$), but interactions with retriever/generator choices may confound effects. We mitigate this by fixing the generator and varying only the memory mechanism.

\textbf{External Validity.} Results on NQ/HotpotQA and a single domain corpus may not generalize to all domains (e.g., code, multilingual). We partially address this by including both single- and multi-hop QA and a specialized corpus.

\textbf{Construct Validity.} Retrieval metrics (NDCG/P/R) and answer metrics (EM/F1) capture complementary aspects of quality; however, they may miss calibration, faithfulness, and latency. We therefore also track cost (retrieval calls/query), memory size, and update overhead.

\textbf{Conclusion Validity.} Multiple hypothesis testing across hyperparameters can inflate chance findings. We report consistent trends across seeds and include conservative configurations recommended for deployment.

\section{Limitations and Future Work}
Future work includes (i) hierarchical multi-tier memories (short/medium/long-term), (ii) learning $\theta,\gamma,\alpha$ from feedback, (iii) task-aware decay schedules, and (iv) multi-modal and graph-augmented dynamic indices.

\section{Conclusion}
Adaptive RAG Memory introduces selective remembrance and decay into the retrieval store, yielding a compact, usage-aligned memory that matches static baselines with fewer parameters. ARM closes the gap between static indices and lifelong, human-inspired memory in practical RAG systems.

\section*{Reproducibility Notes}
The repository includes demos for remembrance and decay analysis, a lightweight benchmark script, and configuration profiles for conservative, balanced, and aggressive adaptation.

\section*{Acknowledgments}
We thank the open-source communities behind FAISS and Sentence-Transformers for foundational tools.

\bibliographystyle{IEEEtran}
\bibliography{refs}

\appendices
\section{Basic Usage (Code)}
\begin{lstlisting}[language=Python,caption={Minimal usage of the Dynamic RAG interface.}]
from dynamic_rag import DynamicRAG

# Initialize the system
rag = DynamicRAG()

# Add documents
rag.add_document("doc1", "Machine learning is a method of data analysis...")
rag.add_document("doc2", "Deep learning uses neural networks...")

# Query the system
results = rag.query("What is machine learning?", top_k=3)

# View memory statistics
stats = rag.embedder.get_memory_statistics()
print(f"Remembered tokens: {stats['remembered_tokens']}")
\end{lstlisting}

\section{Configuration Profiles}
\begin{lstlisting}[language=Python,caption={Conservative vs.\ aggressive adaptation profiles.}]
# Remembrance Parameters
remembrance_threshold = 3   # accesses needed to be 'remembered'
decay_grace_period   = 5    # steps after last access before decay
decay_rate           = 0.95 # multiplicative decay on stale items

# Conservative Remembrance (Recommended)
conservative = dict(remembrance_threshold=5, decay_grace_period=2, decay_rate=0.98)

# Aggressive Adaptation
aggressive = dict(remembrance_threshold=2, decay_grace_period=10, decay_rate=0.80)
\end{lstlisting}

\end{document}